\documentstyle[preprint,aps]{revtex}

\def\hef{ $^4$He }

\def\rr { {\bf r } }
\def\rp { {\bf r' } }

\tighten

\begin{document}

\draft

\title{ Density functional calculations for $^4$He droplets }

\author{ M. Casas$^a$, F. Dalfovo$^b$, A. Lastri$^b$, Ll. Serra$^a$,
and S. Stringari$^b$}

\address{$^a$ Departament de Fisica, Universitat de les Illes Balears \\
E-07071 Palma de Mallorca, Spain }

\address{$^b$ Dipartimento di Fisica, Universit\`a di Trento
I-38050 Povo,  Italy }

\date{\today}
\maketitle

\begin{abstract}
A novel density functional, which accounts correctly for
the equation of state, the static response function and the
phonon-roton dispersion in bulk liquid helium, is used to
predict static and dynamic properties of helium droplets.
The static density profile is found to exhibit significant
oscillations, which are accompanied by deviations of the
evaporation energy from a liquid drop behaviour in the case
of small droplets.  The connection between such oscillations
and the structure of the static response function in the
liquid is explicitly discussed.  The energy and the wave
function of excited states are then calculated in the framework
of time dependent density functional theory. The new functional,
which contains backflow-like effects, is expected to yield
quantitatively correct predictions for the excitation spectrum
also in the roton wave-length range.
\end{abstract}


\narrowtext

\section{Introduction}
\label{sec:intro}

The study of Helium droplets, as a prototype of  finite quantum fluid,
has been the object of extensive experimental and theoretical investigations
in the last decade (see Ref.~\cite{Wha94} for a recent review).  The
ground state and the excited states of pure $^4$He droplets, as well
as of droplets doped with atomic and molecular impurities, have been
studied using different theoretical approaches. However, the overlap
between theories and experiments is not yet satisfactory.

In this paper we investigate static and dynamic properties of $^4$He
droplets using a density functional theory. We use a functional
recently introduced by the Orsay-Trento collaboration \cite{ot94}.
It is an extension of a previous  phenomenological functional \cite{Dup90},
which has been extensively used in the last years to study the static
and the dynamic properties of  inhomogeneous phases of liquid helium
(free surface, films, droplets).  The Orsay-Trento functional improves,
by construction, the  description of relevant  properties of the bulk
liquid, such as the $q$-dependence of the static response function and
the phonon-roton  dispersion. We study here how this improvement affects
the results  for the static and dynamic properties of helium droplets.
In particular, we look for deviations of the density
profile and the evaporation energy from the liquid drop behaviour
and we explore the effects of backflow correlations on the excitation
spectrum at large momentum transfer.

The paper is organized as follows. In Sect.~\ref{sec:dft} we present briefly
the functional (for a complete discussion we refer to \cite{ot94}). In
Sect.~\ref{sec:groundstate} we apply it to the ground state of the droplets.
Then, in Sect.~\ref{sec:tddf},  we sketch the formalism of time  dependent
density functional  theory and present some results for  excited states.

\section{Density Functional}
\label{sec:dft}

A first systematic analysis of the ground state properties
of $^4$He  and $^3$He droplets in the framework of density
functional theory was done in Ref.~\cite{Str87}, using a
simple functional based on a  zero-range  effective
interaction.  The same functional was later  generalized to
include finite-range effects \cite{Dup90}.   This functional has been
applied to investigate static
properties of several inhomogeneous  systems, such as helium films
and wetting phenomena \cite{Pav91}, vortices in bulk liquid
\cite{Dal92} or droplets \cite{Dal94}.

In density functional theory the energy  of the system is
written as
\begin{equation}
E = \int
\! d{\bf r} \  {\cal H} [\Psi,\Psi^*]  \ \ \ ,
\label{e}
\end{equation}
with the complex function $\Psi$ in  the form
\begin{equation}
\Psi({\bf r},t) = \Phi({\bf r},t) \exp \left( {i \over \hbar}
S ({\bf r},t) \right) \ \ \ .
\end{equation}
The real function $\Phi$ is related to the diagonal one-body
density by $\rho=\Phi^2$, while the phase $S$ fixes the velocity
of the fluid through the relation ${\bf v}= (1/m) \nabla S$.
The ground state of the system is found by minimizing the
energy with respect to the density, while the time dependent
theory is formally equivalent to the RPA equations with an effective
particle-hole interaction, as we will see in Sect.~\ref{sec:tddf}.
We use the  phenomenological density  functional of Ref.~\cite{ot94},
containing  a few parameters which are fixed to reproduce bulk
properties of liquid helium. It has the form:
\begin{equation}
E \ = \ E^{(kin)} [\rho,{\bf v}] +  E^{(c)}[\rho] +
E^{(bf)}[\rho,{\bf v}] \; ,
\label{eq:ot}
\end{equation}
where the first term is the kinetic energy of non interacting bosons,
\begin{eqnarray}
E^{(kin)} [\rho,{\bf v}]
        \ &=& \ \int \! d\rr \ {\hbar^2 \over 2m}|\nabla \Psi({\bf r})|^2
\nonumber  \\
            &=& \int\! d{\bf r} \bigg\{
                {\hbar^2 \over 2m} (\nabla \sqrt{\rho})^2
		+ {m \over 2} \rho(\rr) |{\bf v}({\bf r})|^2 \bigg\} \; ,
\label{eq:ekin}
\end{eqnarray}
the correlation energy $E^{(c)}$ is given by
\begin{eqnarray}
E^{(c)}[\rho]  &=& \ \int \! d{\bf r} \Big\{
        {1\over 2} \int \!  d\rp \ \rho(\rr) V_l(|\rr-\rp|) \rho(\rp)
	\ + \ {c_2 \over 2} \rho(\rr)  (\bar \rho_{\rr})^2
        \ + \ {c_3  \over 3} \rho(\rr)  (\bar \rho_{\rr})^3
\nonumber
\\     &-& {\hbar^2 \over 4m} \alpha_s \int \! d\rp \ F(| \rr -\rp |)
	       \left(1-{\rho(\rr) \over \rho_{0s}} \right)
	       \nabla \rho(\rr) \cdot \nabla \rho(\rp)
	       \left(1-{ \rho(\rp) \over \rho_{0s}} \right) \Big\}  \; ,
\label{eq:ec}
\end{eqnarray}
and, finally,  the backflow energy $E^{(bf)}$ is
\begin{equation}
E^{(bf)} [\rho,{\bf v}] \  = \
 - {m \over 4} \int \! d{\bf r} d\rp \ V_J (|\rr-\rp|) \ \rho(\rr)
 \rho(\rp) \  \left[ {\bf v}(\rr) -{\bf v}(\rp) \right]^2 \; .
\label{eq:ebf}
\end{equation}
The first term in the kinetic energy, which depends
on gradient of the density,   is a quantum pressure;  it
corresponds to  the zero temperature kinetic energy of
non-interacting bosons of mass $m$. The two-body interaction $V_l$ in
the correlation energy $E^{(c)}$ is the Lennard-Jones
interatomic potential, with the standard parameters  $\alpha
=2.556$ \AA \ and $\varepsilon=10.22$ K, screened at short
distance ($V\equiv 0$ for $r<h$, with $h=2.1903$\AA). The two
terms with the parameters  $c_2 = -2.411857 \times 10^4$ K \AA$^6$
and $c_3=1.858496 \times 10^6$ K \AA$^9$
account phenomenologically for short range correlations
between atoms. The  weighted density $\bar \rho$ is the
average of $\rho (\rr)$ over a sphere of radius $h$. Those
terms are very similar to the functional of Ref. \cite{Dup90}.
The last term in $E^{(c)}$,  depending on the gradient of the
density in different points, is new and has been added in
order to improve the description of the static response function
in the roton region. The function $F$ is a
simple gaussian, $F (r)= \pi^{-3/2} \ell^{-3} \exp(-r^2/\ell^2)$
with  $\ell=1$~\AA, while  $\alpha_s=54.31$~\AA$^3$ and   $\rho_{0s}
=0.04$ \AA$^{-3}$. The energy $E^{(bf)}$ contains an effective
current-current interaction accounting for backflow-like correlations.
In Ref.~\cite{ot94} the simple parametrization
\begin{equation}
 V_J (r) = (\gamma_{11} + \gamma_{12} r^2) \exp(-\alpha_1 r^2)
         + (\gamma_{21} + \gamma_{22} r^2) \exp(-\alpha_2 r^2) \ \ \ ,
\label{eq:vj}
\end{equation}
was chosen in order to reproduce the phonon-roton dispersion in
bulk liquid. The parameters are given in Table \ref{table}.

A detailed discussion about the meaning of the different contributions
to functional (\ref{eq:ot}) and about the choice of the parameters is
given in Ref.~\cite{ot94}. Here we apply this functional to the statics
and the dynamics of droplets.

\section{Ground state}
\label{sec:groundstate}

To find the ground state of the system one has to minimize
the energy
\begin{equation}
E_0 \ = \ E^{(c)}[\rho] \ + \ \int \! d\rr  \ {\hbar^2 \over 2m}
( \nabla \sqrt{\rho})^2
\label{eq:e0}
\end{equation}
with respect to the particle density $\rho$.
This leads to the Hartree-type equation
\begin{equation}
\left\{ - {\hbar^2 \over 2m} \nabla^2  + U [\rho, \rr] \right\}
\sqrt{\rho(\rr)} \ = \ \mu \sqrt{\rho(\rr)} \ \ \ .
\label{eq:hartree}
\end{equation}
The Hartree potential  is given by $U[\rho, \rr]
\equiv \delta E^{(c)} /\delta \rho(\rr)$. The chemical potential $\mu$
is introduced in order to ensure the  proper normalization  of the
density to a fixed number of particles.

For helium droplets the Hartree equation has to be written
in spherical geometry. It becomes a  unidimensional integro-differential
equation which requires an iterative procedure to be solved. The
resulting density profile, normalized to the bulk value
$\rho_0=0.021836$ \AA$^{-3}$, for small droplets  ($10 \le N
\le 60$) is shown in Fig.~\ref{fig:3d}. One clearly notice the
appearance of oscillations both in the interior of the droplets
and on the surface.  The central density approaches the bulk
density  for large droplets, but it deviates significantly from the smooth
behaviour predicted for a drop of compressible liquid \cite{Str87}.
Such oscillations were not found with previous functionals.

Shell effects in $^3$He droplets, originating from Fermi statistics
and shell closure in the single particle spectrum \cite{Str85},
are expected to be important. On the contrary, the existence of
a shell structure in bosons droplets is still under debate.
Twenty years ago Regge and Rasetti \cite{Reg72}  suggested the
possibility to have \lq\lq magic"  droplets by
looking at the structure of the   elementary excitations in
the bulk. In particular they showed that the surface of the
liquid drop could act  as a source of excitations producing
static ripples on the density profile. In this approach
the static ripples of the surface profile
are connected to the pronounced  peak  of the static response
function of bulk  liquid at the roton wave length.
The authors of Refs. \cite{Reg72}
also commented about the possibility that those
oscillations could be washed out by the zero-point
motion of the surface, whose thickness is expected to be
larger than the interatomic distance.
Indeed  the subsequent theoretical
predictions for the surface profile in helium droplets
\cite{Str87,Pan83,Cep89,Ram90},  predicted a
thickness of the order of $6 \div 8$ \AA\ and no density
oscillations.  A systematic analysis of small droplets with
Green's Function Monte Carlo techniques was presented by
Melzer and Zabolitzky \cite{Mel84} in a work entitled \lq\lq No
magic numbers in neutral \hef clusters". However, the
problem has been re-opened by the most recent Diffusion
Monte Carlo calculations by Chin and Krotscheck \cite{Chi92}
which show clear structures in the density profile of
droplets with 20, 40, 70, and 112 atoms.

In order to check the relation between the density oscillations in
Fig.~\ref{fig:3d} and the behavior of the static response function
$\chi(q)$ we calculate the latter using our density functional theory.
It can be evaluated by taking the second functional derivative of
the energy through the relation
\begin{equation}
- \chi^{-1}(q) = {\hbar^2 q^2 \over 4 m} + {\rho \over V}
\int \! d\rr d\rp { \delta^2 E^{(c)} \over \delta \rho(\rr) \delta
\rho (\rp) } e^{- i {\bf q} (\rr -\rp) } \ \ \ ,
\label{eq:secder}
\end{equation}
where $V$ is the volume occupied by the system.
A major advantage of liquid helium  is that $\chi(q)$ is known
experimentally since it is also equal to the inverse energy moment
of the dynamic structure function $S(q,\omega)$ which is measured in
neutron scattering. In the upper part of Fig.~\ref{fig:chi} the
experimental value  of the static response function at zero
pressure \cite{Cow71} is compared with the predictions of
three functionals: the one of the present work \cite{ot94} and the
ones of Refs.~\cite{Dup90,Str87}. The static response
function is strongly $q$-dependent, showing a peak at
the roton  wave length. According to   the theory of Regge and
Rasetti  this peak is also related to  density oscillations in
the profile of the free surface and  of droplets.  In the lower
part of Fig.~\ref{fig:chi} we show the  predictions for the density
profile of the free surface with the three  functionals. Indeed,
while the three surface profile have a similar thickness, of the
order of 6 to 8 \AA, they have a different structure. Only the
functional of the present work, which reproduce completely the
peak of $\chi(q)$ at the roton wave length, exhibits density
oscillations. We note also  that the same physical effect
can be understood in terms of the static structure factor $S(q)$,
which has also a peak in the roton wavelength region. The relation
between the quantities $\chi(q)$ and $S(q)$ in the framework of the
present density functional theory is discussed in Ref.~\cite{ot94}.
It is worth mentioning that small density oscillations at the liquid-vapour
interface have been recently predicted in classical fluids interacting
through short ranged potentials \cite{Hen92}. In that case, the
connection with the behavior of the radial distribution function,
$g(r)$, has been investigated in detail. For classical fluids the
static response function and the radial distribution function are
related through the Ornstein-Zernike equation.

The oscillations of the free surface profile are very small; they are
not expected  to give rise to measurable effects. The
surface energy is practically the same for the three profiles in
Fig.~\ref{fig:chi} (about $0.27$ K\AA$^{-2}$) and, moreover, the available
experimental data on the surface reflectivity \cite{Lur92} are not
sensitive to such small oscillations in the profile. In the case
of small droplets, however, the oscillating structure of the surface
can combine with finite size effects, i.e.,  the tendency to form
closed shell of atoms. This seems the case for the profile in
Fig.~\ref{fig:3d}. We notice that the structure in the density of
the droplets  is not related to a \lq\lq solid-like"
behaviour; the oscillations originate at the surface, while a
solid structures are not favoured in general by the presence of
surfaces.

The comparison between our results for the density profile and the ones
of {\em ab initio} calculations is given in Fig.~\ref{fig:dmc}
for droplets with $20$ and $70$ particles. The solid
line is the result of the present density functional calculation, while the
DMC results of Ref.~\cite{Chi92} are represented by solid circles.
The DMC data
exhibit more pronounced oscillations, but possible spurious effects of
metastable states, slowing  down the convergence in the  Monte Carlo
algorithm,
cannot be completely ruled out \cite{Chi92}.  Recently the same  authors have
found oscillations in $\rho(r)$ even with a variational  approach based on
the HNC approximation  \cite{Chi95} (dashed line). Even though the HNC method
underestimates the central density, it predicts oscillations with amplitude
and phase in remarkable agreement with the ones of density functional
theory.
An even better agreement is found in the most recent DMC calculations
by Barnett and Whaley \cite{Bar94} (empty circles),
where  the statistical error
is significantly reduced with respect to  Ref.~\cite{Chi92}.

Our predictions for the energy per particle are shown in
Fig.~\ref{fig:en} (solid line), together with the results of
previous Monte Carlo and density functional calculations.
The energy is  a smooth function of the particle number $N$.
Indeed \hef clusters behave \lq\lq grosso modo" as liquid droplets
and the energy can be easily fitted with the  liquid drop formula
\begin{equation}
{E \over N} = a_v + a_s N^{-1/3} + a_c N^{-2/3} + a_0 N^{-1}
\ \ \  ,
\label{eq:drop}
\end{equation}
where the volume coefficient $a_v$ is the chemical potential
in bulk,  the surface coefficient is fixed by the free
surface energy, while $a_c$ and $a_0$ can be taken as fitting
parameters.  The energy calculated with the density functional
differs from the liquid drop fit by less than $0.02$ K  for all
droplets with $N> 30$.  This seems to rule out  apparently any shell
effect in the energy systematics.  However, one must keep
in mind that the most relevant quantity determining the
mass distribution of droplets is the evaporation
energy $[E(N-1)-E(N)]$. The latter is not as smooth as the
energy per particle. In Fig.~\ref{fig:evap} we show the evaporation
energy predicted by the density functional (solid line) and  the
one obtained with the liquid drop formula (\ref{eq:drop}) (dashed
line).  The difference between the two curves is also shown
(dot-dashed line). When the difference is positive
the droplets are more stable than it is predicted by the liquid drop
formula.  We note clear oscillations, having decreasing amplitude
and increasing periodicity as a function of $N$. The same kind of
oscillations appear in the central density of the droplets,
as seen in  Fig.~\ref{fig:3d}. Since the distance between two crests
of the  surface ripples is practically constant  and  the droplet radius
goes  approximately like $N^{1/3}$, the period of oscillations
of the central density as a function of $N$, as well as the one of
the evaporation energy,  increases as $N^{1/3}$.

These results support quantitatively the original pictures proposed
in Ref. \cite{Reg72}, where the surface ripples have been discussed
in terms of {\em soft sphere closed packing}.  However the predicted
deviations of  the evaporation energy from  the liquid drop behaviour
are very  small (less than $0.1$ K).  This makes their detection quite
difficult since the temperature of the droplets in the
experimental beams is presently expected to be about $0.4$ K.

In conclusion, we have shown some evidence for the
occurrence of shell effects in neutral  \hef droplets.
They are predicted by a density functional theory and
physically originate by the strong peak in the  static response
function of liquid helium at the roton wave length. The
shell effects  are visible in the density profiles and the
evaporation energy as a function of the number of particles
in the droplet. Our  results are in agreement with recent
Monte Carlo simulations \cite{Chi92} and seem to confirm
the previous predictions by Regge and Rasetti \cite{Reg72}.

\section{Excited States}
\label{sec:tddf}

\subsection{Density-density response}

The first theoretical investigations of the elementary excitations
of $^4$He droplets have been based on a generalized Feynman theory
within the framework of microscopic Monte Carlo
calculations of the ground state \cite{Pan83,Ram90,Chi90,Chi92}.
A systematic discussion, using a Diffusion Monte Carlo algorithm,
has been presented in \cite{Chi92}. In general, within this
framework, it has been found that the dispersion with transferred
momentum
of the collective energies approaches the Feynman spectrum for
bulk $^4$He as the cluster size increases.   The convergence is very fast
and already at small sizes ($N\approx 70$) there seems to be a well
defined {\em roton minimum}. The Feynman spectrum
is known however  to overestimate by almost a factor two the experimental
roton energy.

Since the Orsay-Trento functional reproduces the bulk phonon-roton
spectrum, it is interesting to explore what it gives for the dynamics of
helium droplets.  In the following we study the behaviour of monopole
($L=0$)
and quadrupole ($L=2$) excitations at low $q$, where the effect
of the velocity dependent term turns out to be negligible, as well as the
behaviour of monopole excitations at large $q$, where the spectrum
is shown to approach the correct phonon-roton dispersion. We discuss
the deviations from the results of a liquid drop model and the comparison
with previous theories.

The formalism of the time dependent density functional theory has
been already introduced, for instance, in Refs.~\cite{Las95,Cas90,ot94}
for applications to the dynamics of the free surface, of droplets,
and films of helium on solid substrates. The same theory is here developed
using the formalism of Green's Functions, allowing for a direct
evaluation of the dynamic response function.
We consider elementary excitations which are induced by an external
field that couples to the particle density in the droplet.
For sufficiently weak external fields the response can be treated linearly
within the Random-Phase approximation. The response function to a
transition operator $Q({\bf r})$ is defined as
\begin{equation}
{\cal R}(\omega) =
\int\! d{\bf r} d{\bf r}'\
Q^\dagger({\bf r}) G({\bf r},{\bf r}';\omega) Q({\bf r})\; ,
\label{es1}
\end{equation}
where $G$ is the retarded Green's function, which is defined in terms of the
ground state ($\vert 0>$), excited states ($\vert n>$) and their
corresponding excitation energies $\omega_{n0}$, and the creation
operator $\psi^\dagger({\bf r})$ (assuming time-reversal
invariance of the matrix elements):
\begin{eqnarray}
G({\bf r},{\bf r}';\omega) = &-&
\sum_n
<0\vert \psi^\dagger({\bf r})\psi({\bf r})\vert n>
<n\vert \psi^\dagger({\bf r}')\psi({\bf r}')\vert 0> \nonumber\\
& & \Big[
{1\over \omega_{n0}-\omega-i\eta} + {1\over \omega_{n0}+\omega+i\eta}
\Big]
\; .
\label{es2}
\end{eqnarray}
{}From Eq.\ (\ref{es2}) it is clear that the excitation energies are at
the poles of $G$, and that they produce sharp peaks in the imaginary part
of ${\cal R}(\omega)$.

Within the RPA the Green's function is calculated from the equation
\begin{equation}
G({\bf r},{\bf r}';\omega) =
G^{(0)}({\bf r},{\bf r}';\omega)+
\int\! d{\bf r}_1 d{\bf r}_2  \ G^{(0)}({\bf r},{\bf r}_1;\omega)
V_{ph}({\bf r}_1, {\bf r}_2)
G({\bf r}_2,{\bf r}';\omega) \; ,
\label{es3}
\end{equation}
where $G^{(0)}$ is the Green's function for a helium droplet with $N$ atoms
within the single-particle model,
i.e., expressed in terms of single particle wave functions $\varphi_i$ and
energies $\varepsilon_i$:
\begin{equation}
G^{(0)}({\bf r},{\bf r}';\omega) = -N
\varphi_0^*({\bf r})\varphi_0({\bf r}')
\sum_n \varphi_n({\bf r})\varphi^*_n({\bf r}') \Big(
{ 1 \over \varepsilon_n-\varepsilon_0-\omega-i\eta } +
{ 1 \over \varepsilon_n-\varepsilon_0+\omega+i\eta} \Big)
\; ,
\label{es4}
\end{equation}
and $V_{ph}$ is the residual (particle-hole) interaction. The sum in
Eq.\ (\ref{es4}) extends, in principle, to all states $\varphi_n$, including
those lying within the energy continuum (see Appendix).
Once this is obtained, the RPA equation can be solved as a matrix
equation in coordinate space and, finally, the response function can be
obtained  from Eq.~(\ref{es1}).

In principle, the external field to be used  in density-density response
is the plane wave, with transferred
momentum ${\bf q}$,
$Q({\bf r}) = \sum_{i=1}^N{ e^{i{\bf q}\cdot{\bf r}} }$.
However, with the multipole expansion indicated in the Appendix,
one can calculate separately
the response to each multipole
$Q_L=\sum_{i=1}^N{ j_L(qr_i) Y_{L0}(\hat{r}_i) }$.
In the limit
of low $q$ this external field reduces to
$Q_L=\sum_{i=1}^N{ r_i^L Y_{L0}(\hat{r}_i) }$.
Within the previous formulation of the response we will include exactly the
coupling with the particle continuum. This may have some importance,
especially
for the response at high momentum transfer.

\subsection{Residual particle-hole Interaction}

The particle-hole interaction entering Eq.\ (\ref{es3}) is obtained, within
density-functional theory, from the second variation of the energy
functional. For functionals depending just on particle density $\rho({\bf r})$
and its gradients it may be obtained in the following way.

Let the density be written, in general, in terms of the single particle
basis $\varphi_i({\bf r})$ as
\begin{equation}
\rho({\bf r}) = \sum_{ij} \rho_{ij}
\varphi^*_i({\bf r})\varphi_j({\bf r}) \; .
\label{ri1}
\end{equation}
The second variation of the correlation energy $E_c$
with respect to $\rho_{ij}$'s, taken at the ground state,
provides the two-body interaction $V_{ph}$:
\begin{equation}
<ij\vert V_{ph}({\bf r}_1,{\bf r}_2)\vert kl> =
{\delta^2 E^{(c)}[\rho]\over \delta\rho_{ik} \delta\rho_{jl} }\; .
\label{ri2}
\end{equation}
The variation in this equation may be calculated as
\begin{eqnarray}
{\delta^2 E^{(c)}[\rho]\over \delta\rho_{ik} \delta\rho_{jl} }&=&
\int d^3r_1 d^3r_2
\left(
{\delta^2 E^{(c)}[\rho]\over \delta\rho({\bf r}_2) \delta\rho({\bf r}_1)}
\right)_{g.s.}
{\delta\rho({\bf r}_1)\over\delta\rho_{ik}}
{\delta\rho({\bf r}_2)\over\delta\rho_{jl}}\nonumber\\
&=& \int d^3r_1 d^3r_2
\varphi_i^*({\bf r}_1) \varphi^*_j({\bf r}_2)
\left(
{\delta^2 E^{(c)}[\rho]\over \delta\rho({\bf r}_2) \delta\rho({\bf r}_1)}
\right)_{g.s.}
\varphi_k({\bf r}_1) \varphi_l({\bf r}_2) \; ,
\label{ri3}
\end{eqnarray}
and, consequently, we obtain the well known result
\begin{equation}
V_{ph}({\bf r}_1,{\bf r}_2) =
\left(
{\delta^2 E^{(c)}[\rho]\over \delta\rho({\bf r}_2) \delta\rho({\bf r}_1)}
\right)_{g.s.} \; .
\label{ri4}
\end{equation}
The Fourier transform of this quantity enters the definition of the static
response function $\chi(q)$ in bulk liquid, as  in Eq.~(\ref{eq:secder}).
{}From a practical viewpoint, the functional derivative implies the calculation
of derivatives of the integrands in Eq.~(\ref{eq:ec}) with respect
to $\rho$ and $\nabla \rho$.

The inclusion of the backflow term in the functional requires a
generalization of the previous scheme. This term introduces a dependence
on the current  density ${\bf j}$, defined as
\begin{equation}
{\bf j}({\bf r}) =
{\hbar\over i2m} \sum_{ij} \left. (\nabla-\nabla')\rho_{ij}
\varphi_i^*({\bf r}) \varphi_j({\bf r}') \right\vert_{{\bf r}={\bf r}'}
\; .
\label{ri6}
\end{equation}
Since the current density vanishes for the ground state, the only term
contributing to the residual (backflow) interaction will be
\begin{eqnarray}
{\delta^2 E^{(bf)}[\rho,{\bf j}]\over \delta\rho_{ik} \delta\rho_{jl} }&=&
\int d^3r_1 d^3r_2
\left(
{\delta^2 E^{(bf)}[\rho,{\bf j}]\over
\delta{\bf j}_\nu({\bf r}_2) \delta{\bf j}_\nu({\bf r}_1)}
\right)_{g.s.}
{\delta{\bf j}_\nu({\bf r}_1)\over\delta\rho_{ik}}
{\delta{\bf j}_\nu({\bf r}_2)\over\delta\rho_{jl}} \; ;
\label{ri7}
\end{eqnarray}
where the index $\nu$ spans the three components, and a sum over $\nu$
is assumed. From this expression one finds for the back-flow part of
the residual interaction
\begin{equation}
V^{(bf)}_{ph}(1,2) =
-{\hbar^2\over 4m^2}
\left(
{\delta^2 E^{(bf)}[\rho,{\bf j}]\over
\delta{\bf j}_\nu({\bf r}_2) \delta{\bf j}_\nu({\bf r}_1)}
\right)_{g.s.}
(\roarrow\nabla_1-\loarrow\nabla_1)
(\roarrow\nabla_2-\loarrow\nabla_2) \; ,
\label{ri8}
\end{equation}
where the parenthesis is in fact independent of $\nu$ and the
gradients act only on the single-particle wave functions.
This last expression
shows explicitly how the current-dependent term yields a velocity-
dependent
residual interaction. Indeed, Eq.~(\ref{ri8}) is quite similar to
the Skyrme residual interaction in nuclear physics \cite{BT75}, and we
will use  a similar technique to calculate the associated RPA response.
Summarizing, we have two contributions to the residual interaction
\begin{equation}
V_{ph} = V^{(0)}_{ph} + V^{(bf)}_{ph}\; ,
\label{ri19}
\end{equation}
where $V^{(0)}_{ph}$ is the residual interaction, obtained with
Eq.\ (\ref{ri4}), associated to the functional without current terms,
while $V^{(bf)}_{ph}$ is the contribution from these current terms.

\subsection{Results}

In this section we present the results obtained for
${\cal R}(\omega)$ in the formalism presented above. We use the
technique developed in Ref.\ \cite{BT75,LG80} and already applied to
$^4$He droplets in the context of a contact (zero-range) effective
interaction in Ref.\ \cite{Cas90}. The present formalism generalizes
that of \cite{Cas90} by including both finite range and back-flow
effects. We expect these to be important for the droplet response
at high transferred momentum $q$, while for low $q$ we expect to
recover a surface-mode systematics similar to that of
Ref.\ \cite{Cas90}. The present functional reproduces by construction
the dispersion of the elementary excitations in bulk liquid
$^4$He \cite{ot94},  i.e., the phonon-roton curve up to
$q\approx 2.3$~\AA$^{-1}$. Consequently, it is particularly appropriate
for the study of the response of droplets in this region of $q$'s.

\subsubsection{Low $q$ results}

Fig.~\ref{fig:d1} shows the imaginary part of ${\cal R}(\omega)$ for
the low $q$ limit of the monopole and quadrupole external fields, which are
$Q=\sum_i{r_i^2}$ and $Q=\sum_i{r_i^2Y_{20}}$, respectively,
for the droplet with $N$=112 atoms.
The monopole corresponds to a {\em breathing} mode, while the quadrupole
is a {\em surface vibration} \cite{Cas90}. It is clearly seen how a
very intense peak appears in the RPA result (continuous line),
corresponding to a collective oscillation of the $^4$He atoms.
The independent-particle result (dashed line), obtained by neglecting in
Eq.\ (\ref{es3}) the particle-hole interaction, is in both cases
more fragmented, revealing that the RPA correlations play a crucial role
in the dynamics of the system.

The effect of the current dependent part of the residual interaction,
$V_{ph}^{(bf)}$, on the monopole and quadrupole excitations in this
range of $q$ is completely negligible. We have explicitly checked that
the monopole result in Fig.~\ref{fig:d1} is not affected at all by
$V_{ph}^{(bf)}$. This is consistent with the fact that in bulk liquid
the lowest excitation for $q \to 0$  is the phonon-mode, which is not
affected by backflow-like correlations. As concern the $L \ne 0$
modes of the clusters, Chin and Krotscheck \cite{Chi95} have recently
shown that, at low $q$, they  approach the dispersion of ripplons
on  the free surface. The latter, again, is found to be practically
unaffected by backflow correlations \cite{Las95}.

The static induced densities $\delta\rho^{(0)}(r)$ for the same cluster
are plotted in Fig.~\ref{fig:d2}.
This figure proves the behaviour mentioned before:
the monopole induced density has a node and penetrates the interior of
the cluster, while the quadrupole is almost completely localized at the
droplet surface. An interesting result shown by this figure is the appearance
of small oscillations in $\delta\rho^{(0)}$, in the inner part, both
for the monopole and the quadrupole. This behaviour is connected to the
oscillations displayed by the droplet equilibrium density $\rho$, and is due
to the repulsive core of the effective interaction
(see also Ref.~\cite{BH94}).  In fact, with the use of a contact
interaction the small oscillations disappear both in $\rho$
and $\delta\rho$. Similar oscillations have also been found
using other microscopic methods, like  Monte Carlo calculations \cite{Chi92}.

The dependence of the collective energy with the number of atoms in the
cluster
is shown in Fig.~\ref{fig:d3}. The comparison with independent particle
predictions shows that the role of RPA correlations increases with $N$.
The deviations of the RPA predictions from the Liquid Drop Model (LDM)
results
(long-dashed line, see \cite{Cas90}) are due to finite-size effects and are
important up to $N \simeq$ 100 for surface vibrations and up to $N \simeq$
500
for compression modes.
The energies are in nice agreement with those of
Ref.~\cite{Cas90}, confirming the expectation that the response at
$q\to 0$ is not affected by the long range part of the atom-atom interaction.
In the following section we show how this is completely different for
the high $q$ region of the response.

\subsubsection{High $q$ results}

In this section we will focus on the $L=0$ response. In this
case, the hole and particle states which contribute to the Green's function
(\ref{es4}) have spherical symmetry and this allows us to take only the
radial
part of the $(\roarrow\nabla-\loarrow\nabla)$ in $V_{ph}^{(bf)}$,
as a consequence, only the monopole term in the multipole expansion
of the second functional derivative of Eq.~(\ref{ri8}) is needed.
The situation is different for higher multipoles, since then the
residual interaction $V_{ph}^{(bf)}$ involves different multipoles
of the second functional derivative of Eq.~(\ref{ri8}). This
greatly complicates the numerical calculation and we leave this case for a
future investigation. In this work we concentrate on the monopole response
to show the influence of the current term in the high-$q$ response of
droplets.

Figure \ref{fig:d4}  shows the contour lines of the surface
${\rm Im}[{\cal R}(q,\omega)]$
for the $N=112$ droplet. One clearly appreciates that the zone of higher
response strength, where the peaks are located, resembles the dispersion
curve of the bulk $^4$He elementary excitations. The existence of a
{\em roton minimum} is clearly seen around $q=2$~\AA$^{-1}$. However,
the
energy of this excitation is quite affected by the current term
$V_{ph}^{(bf)}$. The upper panel is the result when $V_{ph}^{(bf)}$
is not included and provides a minimum energy at $\approx 15$~K.
When this current term is included the minimum energy moves to
$\approx 9$~K. Fig.~\ref{fig:d5} shows how this picture changes for different
cluster sizes. For $N=20$ the number of atoms is not enough to
develop the roton minimum while for $N=728$ the result follows quite
closely
the bulk one. Similar behaviour has been found using other microscopic
methods
\cite{Ram90,Chi92}; being  based on Feynman-like wave functions, those
calculations yield a roton minimum at approximately twice the experimental
roton gap in bulk liquid.

\section{Conclusions}

We have presented static and dynamic  calculations for pure $^4$He droplets
using a new density functional theory developed by the Orsay-Trento
collaboration \cite{ot94}. The density functional is written in such a way
that relevant properties of the uniform liquid (equation of state, static
response function, phonon-roton dispersion) are accurately reproduced. The
theory is suitable to study properties of non uniform states of liquid
helium.

The static density profile and the energy of
helium droplets have been calculated as a function of the number
of atoms. We have found small density oscillations associated with regular
deviations of the evaporation energy from the smooth liquid drop behaviour.
We have discussed these oscillations in connection with the structure
of the static response function, as suggested by a previous model of
Regge and Rasetti \cite{Reg72}. Our results compare well with recent
{\em ab initio} Monte Carlo calculations.

We have also calculated the energy and transition densities of monopole
and quadrupole excited states. The transition densities at $q\to 0$
display small oscillations similar to those of the ground state density.
As in the case of the zero-range interaction, the solution of the RPA
equations reveals the importance of the finite size effects and the
long range correlations in the determination of the excitation energies
of the compression and surface modes. In particular, the deviation from
the LDM results are important up to $N\approx 500$ for compression
modes and up to $N\approx 100$ for surface vibrations.

The analysis of the monopole mode at high momentum transfer shows the
existence of a roton minimum around $q\approx 2$~\AA$^{-1}$ for clusters
with a number of atoms  $N\ge 20$. As in the uniform liquid, the backflow
correlations, which are included phenomenologically in the theory,
yield a sizeable decrease in the energy of this minimum, from
$\approx$15~K
to $\approx$9~K.

\acknowledgements

This work was partially supported by  European
Community grant ERBCHRXCT920075; by INFN, gruppo collegato di Trento;
and by DGICYT (Spain) grant PB92-0021-C02-02.

\appendix
\section{RPA in the continuum}

The sum in Eq.~(\ref{es4}) can be written in a more compact way, using
the particle Green's function $g_p$ corresponding to the particle
Hamiltonian $H_p$, in coordinate space:
\begin{equation}
g_p({\bf r}_1,{\bf r}_2;\omega) =
<{\bf r}_1 \vert {1\over H_{p}-\omega}\vert {\bf r}_2> \; ,
\label{es5}
\end{equation}
as
\begin{equation}
G^{(0)}({\bf r},{\bf r}';\omega) = -N
\varphi_0^*({\bf r}) \varphi_0({\bf r}') \Big(
g({\bf r},{\bf r}';\varepsilon_0+\omega+i\eta) +
g({\bf r},{\bf r}';\varepsilon_0-\omega-i\eta) \Big) \; .
\label{es6}
\end{equation}
The particle Hamiltonian $H_{p}$ is defined from Eq.~(\ref{eq:hartree})
\begin{equation}
H_p = - {\hbar^2 \over 2m} \nabla^2 + U [\rho, \bf r] \; ,
\label{es7}
\end{equation}
with $\rho$ the ground state density.

For spherical droplets each of the functions in Eq.\ (\ref{es3}) can be
expanded in multipoles as
\begin{equation}
G({\bf r},{\bf r}') = \sum_{L M}
G_L(r,r') Y^*_{LM}(\hat{r}) Y_{LM}(\hat{r}') \; ,
\label{es8}
\end{equation}
and, as a result, Eq.\ (\ref{es3}) also separates for different multipoles.
In a symbolic notation
\begin{equation}
G_L=G^{(0)}_L+G^{(0)}_L V_{ph,L} G_L \; .
\label{es9}
\end{equation}

The particle Green's function for each partial wave $g_L$ is
most easily calculated by means of the scattering solutions to the
effective radial Hamiltonian $H_{p,L}$ at energy $\omega$:
\begin{equation}
H_{p,L}u_{L}(r) = \omega u_{L}\; .
\label{es10}
\end{equation}
It is
\begin{equation}
g_L(r,r';\omega) = {2m\over\hbar^2} {u_L(r_<) w_L(r_>)\over rr'W(u_L,w_L)}
\; ,
\label{es11}
\end{equation}
where $u$ and $w$ are, respectively, the regular and irregular at the origin
radial solutions of (\ref{es10}), and $W(u_L,w_L)$ is their Wronskian.
In this way we can obtain $G^{(0)}$ without having to impose any truncation
of
the continuum.

%

%
%

\begin{figure}
\caption{ Density  profile of  \hef  droplets for $10 \le N \le 60$,
normalized to the bulk value.  }
\label{fig:3d}
\end{figure}

\begin{figure}
\caption{ Static response function of liquid  helium
(above) and free surface profile (below). Circles:
experimental data \protect \cite{Cow71}; lines: density functional
calculations with functional of Ref.  \protect\cite{Str87}
(dotted), Ref.  \protect\cite{Dup90} (dashed) and the one
of Eq. (2) (solid). }
\label{fig:chi}
\end{figure}

\begin{figure}
\caption{ Density  profile  of two droplets.  Solid line:
present work; solid circles: DMC calculations of Ref.~\protect \cite{Chi92};
empty circles: DMC calculations of Ref.~\protect \cite{Bar94}; dashed line:
HNC calculations of Ref.~\protect \cite{Chi95}.  }
\label{fig:dmc}
\end{figure}

\begin{figure}
\caption{Energy per particle versus $N$. Solid line:
present work; dotted line: Ref. \protect \cite{Str87}; dashed
line: results with functional of Ref. \protect \cite{Dup90};
empty circles: Ref. \protect \cite{Mel84};
solid circles: Ref. \protect \cite{Chi92}. }
\label{fig:en}
\end{figure}

\begin{figure}
\caption{ Evaporation energy. Solid line: present work; dashed
line: liquid drop formula.  Dot-dashed line: deviation from the liquid
drop formula (axis on the right). }
\label{fig:evap}
\end{figure}

\begin{figure}
\caption{Imaginary part of the response function at
zero momentum transfer ($q=0$) of the droplet with $N=112$ atoms.
Part (a) corresponds to the monopole case for which
the external field is $Q=\sum_i r_i^2$,
and (b) to the quadrupole, with $Q=\sum_i r_i^2 Y_{20}(\hat{r}_i)$.
See text.}
\label{fig:d1}
\end{figure}

\begin{figure}
\caption{Induced densities for the droplet $N=112$ in the static limit,
$\omega=0$, corresponding to the results of Fig.~\protect \ref{fig:d1}.
We used arbitrary vertical scale.
The dashed line shows the ground state density of the same droplet
with the labeled vertical scale. See text.}
\label{fig:d2}
\end{figure}

\begin{figure}
\caption{Energy of the collective peak at $q=0$
as a function of size for the quadrupole and monopole.
The short-dashed line shows, for comparison, the results of
Ref. \protect \cite{Cas90}. The long-dashed line corresponds to the LDM
prediction. }
\label{fig:d3}
\end{figure}

\begin{figure}
\caption{Contour lines of the surface
${\rm Im}[{\cal R}(q,\omega)]$ for the monopole excitation of the
droplet $N=112$. Part (a) is without including the current-dependent
term $V_{ph}^{(bf)}$ while (b) corresponds to the complete functional.
See text.}
\label{fig:d4}
\end{figure}

\begin{figure}
\caption{Same as Fig.~9b, with the complete functional,
for the droplets with $N=20$ (a) and $N=728$ (b).}
\label{fig:d5}
\end{figure}

\begin{table}
\caption{ Values of the parameters used in $V_J(\rr)$, see Eq. (\protect
\ref{eq:vj}).}
\begin{tabular}{c|c|c|c|c|c}
$\gamma_{11}$&$\gamma_{21}$&$\gamma_{12}$&$\gamma_{22}$&$\alpha_1$&$\alpha_2$
\\
\tableline
$-19.7544$&$-0.2395$&$12.5616$ \AA$^{-2}$&$0.0312$ \AA$^{-2}$
&$1.023$ \AA$^{-2}$&$0.14912$ \AA$^{-2}$\\
\end{tabular}
\label{table}
\end{table}

\end{document}